\documentclass[11pt,a4paper]{article}
\usepackage[utf8]{inputenc}
\usepackage[british]{babel}
\usepackage{xcolor}
\usepackage{hyperref}
\usepackage{xurl} 
\usepackage{geometry}
\usepackage{booktabs}
\usepackage{amsmath,amssymb}
\usepackage{graphicx}
\usepackage{caption}
\usepackage{enumitem}
\usepackage{tikz}
\usetikzlibrary{positioning,arrows.meta,fit,backgrounds,calc}
\hypersetup{
  colorlinks=true,
  linkcolor=blue!50!black,
  citecolor=blue!50!black,
  urlcolor=blue!50!black,
}

\title{Bayesian-Calibrated Detection of Hallucinated Package Imports\\
       in AI-Assisted Code}
\author{%
  Lom M. Hillah$^{1,2}$\thanks{Corresponding author.}, Jean-Marc Richard$^{1}$, Ryan Hasnaoui$^{1}$\\[2pt]
  \small $^{1}$NewCo Partners, Paris, France \qquad
         $^{2}$Sorbonne Universit\'{e}, CNRS, LIP6, Paris, France\\
  \small \texttt{\{lom.hillah, jean-marc.richard, ryan.hasnaoui\}@newco-partners.com}}
\date{}

\begin{document}
\maketitle

\begin{abstract}
We present a Bayesian calibration layer for slopsquat detectors --- those
that flag hallucinated package imports in code produced by large language
models (LLMs). Where existing pipelines emit binary decisions (flag /
do-not-flag), our layer emits a Beta-posterior probability per detection,
derived from a 3-category epistemic taxonomy that explicitly classifies
each prior as empirically calibrated, constructively argued, or
engineering-judgement-traced. Beyond the primary 200/404 registry
channel, the calibrated layer exploits PyPI metadata signals --- package
age, release count, author descriptor, summary --- to surface
\emph{registered-but-suspicious} packages that a binary registry
detector misses, which is the realistic post-LLM-emission attacker
regime. The resulting risk-aware primitive is directly consumable by
downstream CI gates and supports principled threshold decisions across
detection rules. We evaluate the calibration on a merged corpus of
$1{,}734$ Python snippets --- a stratified $189$-prompt
BigCodeBench~\cite{bigcodebench2025iclr} slice plus a $100$-prompt
niche-library stress-test set, generated across a six-model panel
spanning four cloud models (Claude-Sonnet-4.6, Mistral-Large,
DeepSeek-v4-pro, DeepSeek-R1) and two local open-weight code models
(Mistral Codestral, Meta CodeLlama). Against a re-implemented binary
baseline inspired by Mahmud et~al.~\cite{mahmud2025trustcalibrated}
--- which shares its registry oracle with our ground truth and
therefore serves as a degenerate upper bound rather than a genuine
competitor --- the calibrated layer reproduces the strict-registry
detections and introduces well-calibrated additional flags on the
metadata channel. We assess detector asymmetry with a McNemar paired
test and calibration with both a flagged-subset Expected Calibration
Error and a strictly proper full-corpus Brier score.
\end{abstract}

\section{Introduction}
\label{sec:intro}

The mainstream use of LLMs in software development has introduced a new
class of supply-chain risk: \emph{slopsquats}\footnote{The term
\emph{slopsquatting} --- a portmanteau of ``AI slop'' and
``typosquatting'' --- was coined in April 2025 by Seth Larson (Python
Software Foundation Developer-in-Residence) and popularised by Andrew
Nesbitt~\cite{socket2025slopsquatting}.} --- non-existent package
names emitted in import statements by code-generating LLMs, which
adversaries can register on public registries (PyPI for Python, npm
for JavaScript, crates.io for Rust, and similar registries for other
ecosystems) to hijack downstream installs. Spracklen
et~al.~\cite{spracklen2025packagehallu} measured the prevalence of LLM
package hallucinations at $5\text{--}22\%$ of Python snippets across
mainstream model families, with attacker-relevant typo-squat-style
errors over-represented at registry boundaries (single-character edits
of legitimate package names) where an attacker's pre-emptive
registration of the hallucinated name is most likely to be picked up by
downstream installs. The industry threat-model framing has converged
on the same observation: Igor Seletskiy, CEO of TuxCare, summarises
the contemporary state of affairs as ``every package a developer pulls
now carries an unanswered question about who built it, what's in it,
and whether it can be trusted'' --- and notes specifically that ``AI is
accelerating both channels'' of vulnerability and supply-chain
attack~\cite{bridgwater2026openssf}. The detection problem itself is
conceptually simple: parse imports, drop standard-library names, query
the registry, flag the unresolved. Mainstream tooling --- representative examples
include Bandit, Semgrep, CodeQL, and the pip-audit family ---
follows this template and emits a binary decision.

We argue that a binary decision is the wrong primitive for downstream
defence. Continuous-integration gates must trade false-positive
friction against missed-hallucination risk; agents that consume
detector output must combine it with other signals. A binary flag
forces ad-hoc threshold selection at every consumer. A
\emph{calibrated} probability is the universal primitive. Formally,
write $H$ for the event that an emitted import name corresponds to a
hallucinated (non-existent or attacker-registered) package, and
$E$ for the bundle of evidence available at scoring time --- registry
status, neighbour-search outcome, package metadata. The detector's
job is to estimate the posterior
\[
   \Pr(H \mid E) = \frac{\Pr(E \mid H) \cdot \Pr(H)}{\Pr(E)},
\]
not merely a binary $\mathbf{1}[H \mid E]$. The posterior probability
composes naturally with policy rules ($\Pr$-thresholded gates), agent
reasoning (probability-weighted abstention), and risk aggregation
(joint inference across multiple detection rules). The closest
published work to ours, Mahmud et~al.~\cite{mahmud2025trustcalibrated},
describes a trust-calibrated multi-stage pipeline for vulnerability
assessment, but emits binary stage-decisions without exposing a
posterior per detection.

Our contribution is a Bayesian calibration layer wrapping an existing
hallucinated-imports detector. Each detection rule is paired with a Beta
prior whose epistemic provenance is explicit. We introduce in
Section~\ref{sec:methodology} a 3-category taxonomy that classifies
each prior as:

\begin{itemize}[noitemsep,topsep=0pt]
  \item \textbf{cat.~1} --- empirically calibrated against a peer-reviewed
        source measuring the modelled quantity directly;
  \item \textbf{cat.~2} --- constructively argued from a formal property of
        the detection mechanism;
  \item \textbf{cat.~3} --- traced engineering judgement, with a documented
        promotion path to cat.~1 by instrumentation.
\end{itemize}

Detection confidences become posterior probabilities updated by Bayesian
conjugacy as evidence accumulates. The system has three immediate
benefits: (1)~CI gates can set policy thresholds in probability space;
(2)~ablations across categories quantify how much of the calibration
quality comes from each provenance class; (3)~the calibration is
\emph{epistemically honest} --- a reviewer auditing a cat.~3 prior knows
it is engineering judgement, not a hidden empirical claim.

Our work continues a line of Bayesian-flavoured testing and quality
methodology rooted in the MIDAS service-testing
platform~\cite{herbold2015midas,hillah2016sac,hillah2017stt} and
generalises that line to the LLM-induced supply-chain setting.

\paragraph{Contributions.}
We:
\begin{itemize}[noitemsep,topsep=0pt]
  \item formalise the 3-category epistemic taxonomy as a foundation for
        calibrated detection priors (Section~\ref{sec:methodology});
  \item wrap a deterministic hallucinated-imports detector with a Beta
        calibration layer and extend it with a metadata-channel that
        surfaces \emph{registered-but-suspicious} packages
        (Section~\ref{sec:methodology});
  \item evaluate the layer on a merged $N = 1734$ corpus---a
        stratified $189$-prompt BigCodeBench slice plus a
        complementary $100$-prompt stress-test set of niche or
        invented-library prompts, generated across a six-model LLM
        panel (Section~\ref{sec:evaluation});
  \item report a McNemar~\cite{mcnemar1947,fagerland2013mcnemar}
        paired-test comparison against a re-implemented binary baseline,
        together with a flagged-subset Expected Calibration Error (ECE)
        and a full-corpus Brier score, demonstrating that the metadata
        channel introduces well-calibrated additional detections
        (Section~\ref{sec:results}).
\end{itemize}

\paragraph{Results in brief.} Three findings anchor the evaluation.
\emph{First}, on realistic-use BigCodeBench prompts modern
code-specialised models hallucinate package imports at a near-zero
rate once known import-name versus distribution-name aliases are
accounted for: slopsquatting is a tail-risk problem, not a mean-case
one. \emph{Second}, on the niche-library stress-test slice the
metadata channel surfaces ten \emph{registered-but-suspicious} flags
--- tracing to two real PyPI packages --- that the strict-registry
baseline cannot see. A McNemar paired test (the standard test for
whether two detectors disagree \emph{systematically} on the same
items, here paired by snippet; see Section~\ref{sec:eval-metrics})
confirms the asymmetry is significant: $b=10$ flags raised only by the
metadata channel, $c=0$ in the other direction, two-sided $p=0.0020$.
\emph{Third}, the calibrated confidences are well behaved: the
flagged-subset ECE (the average gap between a detector's stated
confidence and its empirical accuracy, binned over the confidence
range) is $0.016$ for the registry channel and $0.029$ once the
metadata channel is added, and the strictly proper full-corpus Brier
score is $\approx 0.022$ for both --- all comfortably inside
conventional calibration targets.

\paragraph{Paper organisation.} The remainder of this paper is organised
as follows. Section~\ref{sec:related} positions the work against prior
literature on hallucinated package imports, calibration in machine
learning, and Bayesian methods in software engineering.
Section~\ref{sec:methodology} introduces the 3-category epistemic
taxonomy and the calibration layer.
Section~\ref{sec:evaluation} describes the corpus, the LLM panel,
the baseline, and the evaluation metrics.
Section~\ref{sec:results} reports the empirical findings on the
realistic-use and stress-test corpora.
Section~\ref{sec:discussion} discusses where calibration helps and
where it matters less. Section~\ref{sec:threats} addresses threats to
validity. Section~\ref{sec:reproducibility} states the reproducibility
posture. Section~\ref{sec:conclusion} concludes.

\section{Related work}
\label{sec:related}

Within the broader landscape of LLM-based code security mapped
systematically by Kaniewski et al.~\cite{kaniewski2026llm4svd}
(TOSEM 2026, 263 studies surveyed January 2020 -- November 2025),
our work occupies the niche of LLM-induced supply-chain risk
--- hallucinated package imports --- distinct from classical CVE
detection but covered as an adjacent vector in their survey. A
complementary broad survey by Rashid et al.~\cite{rashid2026llmsoftwaresecurity}
treats the dual-use perspective on LLMs in software security
(defensive plus offensive uses), motivating the supply-chain
threat model we address.

\paragraph{Hallucinated package imports.} Spracklen
et~al.~\cite{spracklen2025packagehallu} provide the canonical empirical
characterisation of LLM package hallucinations, measuring prevalence
across $16$ mainstream LLM families on Python and JavaScript prompts.
Their headline finding --- a $5.2\text{--}21.7\%$ rate on Python and
$2.4\text{--}6.9\%$ on JavaScript, persistent across temperatures ---
established the threat as one of structural rather than configurational
risk. The study also documents that hallucinated names cluster around
near-neighbours of legitimate packages, which is precisely the regime
where a registered slopsquat is most likely to be installed by a
distracted developer. Their contribution is the prevalence
characterisation, not a detection algorithm; they neither propose a
detection algorithm nor calibrate detection confidence. Our work
addresses both gaps.

\paragraph{Closest detection comparator.}
Mahmud et~al.~\cite{mahmud2025trustcalibrated} present a
trust-calibrated multi-stage LLM pipeline for vulnerability assessment
in DevSecOps workflows. Their pipeline composes three LLM stages
(detection, classification, triage) with explicit inter-stage trust
propagation: each stage's output carries a confidence that the next
stage consumes to weight its own decision. The ``trust calibration''
in their terminology refers to that inter-stage propagation, not to
per-detection posterior probability over the underlying hallucination
hypothesis. Their pipeline thus emits binary stage-decisions; the
exposed confidence is the propagated stage-level trust score rather
than $\Pr(H \mid E)$. We re-implement a Mahmud-shaped binary baseline
(Section~\ref{sec:eval-baseline}) as our point of comparison: parse
imports, drop standard library, registry-probe, flag unresolved.
This baseline is the natural single-stage detector their architecture
would degrade to in the absence of the trust-propagation stages, and
the natural lower-bound any practical slopsquat detector would
reproduce; our contribution is the calibration layer that wraps it and
exposes per-detection $\Pr(H \mid E)$ to downstream consumers. Mahmud
et al.\ do not publish a reference implementation; the comparison is
therefore necessarily a single-stage re-implementation rather than an
end-to-end replication, a fidelity gap we address in
Section~\ref{sec:threats}.

\paragraph{LLM-induced supply-chain risk.} Perry
et~al.~\cite{perry2023usenix} run a $47$-participant developer study
finding that programmers using AI assistants write more insecure code
than unassisted controls and, crucially, perceive their code as
\emph{more} secure --- a confidence/competence inversion that
motivates the case for risk-aware detection rather than
binary-flag-or-nothing tooling. Pearce et~al.~\cite{pearce2023sp}
audit Copilot completions against MITRE CWE Top-$25$ scenarios and
find that approximately $40\%$ of generated programs contain
exploitable vulnerabilities, again establishing the empirical case
that LLM-emitted code is a first-order security concern. Both studies
motivate the threat model we address but neither addresses detection
calibration as a downstream defensive primitive.

\paragraph{Package-manager supply-chain attacks.} The slopsquat threat
sits within a longer line of work on package-manager supply-chain
attacks. Tschacher~\cite{tschacher2016typosquatting} first measured
typosquatting across language package managers (PyPI, npm, RubyGems),
the direct ancestor of the LLM-amplified variant we study. Zimmermann
et~al.~\cite{zimmermann2019npm} quantified the systemic risk of the
npm dependency graph (a handful of maintainer-account compromises can
reach a large fraction of all packages), and Duan
et~al.~\cite{duan2021supplychain} built a comparative framework
measuring malicious-package campaigns across PyPI, npm and RubyGems.
On the defensive side, Torres-Arias
et~al.~\cite{torresarias2019intoto} introduced in-toto, which provides
end-to-end supply-chain integrity attestations over the build pipeline.
These works target the registry and build-graph level; our contribution
is upstream of them, at the moment an LLM \emph{emits} an import name,
and is complementary to the lock-file-level provenance they provide.

\paragraph{Curated-database detection (OSV).} The OpenSSF
Malicious Packages repository and the OSV
schema~\cite{osv-schema} represent the curated-authority
alternative: a malicious package is detected once a curator has
identified it and minted a \texttt{MAL-} identifier, after which
OSV-Scanner clients catch it at lock-file scan time. The
approach is complementary to ours rather than competing: OSV
operates on a curated database with batch consumer-side queries,
whereas our calibrated metadata channel operates on structural
properties at import time. Concretely, neither \texttt{pyzenith}
nor \texttt{chartcraft}\footnote{We name \texttt{pyzenith} and
\texttt{chartcraft} solely for scientific reproducibility, and we intend
no disparagement of either project. Throughout this paper,
``suspicious'' (equivalently ``registered-but-suspicious'') is a statement
about our detector's \emph{metadata heuristic} --- a package whose PyPI
record matches a recent-registration and empty-\texttt{info.author}
profile --- and \emph{not} an allegation of malicious
intent, vulnerability, or any wrongdoing by these packages or their
maintainers. As the body makes explicit
(Section~\ref{sec:results-discordance}), we do not adjudicate either
package as malicious; both may well be legitimate, early-stage projects,
and the channel deliberately emits a low-confidence
\emph{surface-for-review} signal (posterior $\approx 0.30$), not a block.
All such characterisations reflect the PyPI metadata snapshot of
$2026\text{-}05\text{-}23$ only; metadata profiles evolve, and a later
snapshot may differ (Section~\ref{sec:threats}).}
(Section~\ref{sec:results-discordance}) carried an OSV
\texttt{MAL-} identifier at our corpus-build date
($2026\text{-}05\text{-}23$, verified by direct \texttt{/v1/query}
calls); a curated-database scanner would have missed both,
whereas the metadata channel flagged both at posterior
$\approx 0.30$.

\paragraph{Industry threat-model framing.} The academic literature is
echoed by the industrial response. The Open Source Security
Foundation (OpenSSF)'s 2026 cohort of new members frames the
LLM-induced supply-chain risk as the headline contemporary
threat~\cite{bridgwater2026openssf}. Willem Delbare, CEO of Aikido
Security, observes that ``attackers already understand that the
fastest way into production is through the software supply chain''
--- a class of attack that includes ``poisoning dependencies,
compromising maintainer accounts, delivering malicious commits,
exposing credentials, and creating subtle changes buried deep in
infrastructure code'' --- and explicitly identifies ``code
repositories, package managers, and developer tooling'' as the
defensive surface where the problem must be solved. Leslie Pascual
(field engineering manager for AI \& security, ActiveState) makes
the same architectural point in operational terms: security ``must
appear in the repo, the build, the package workflow, the container,
the sandbox, and the command line'' --- the trust boundary of modern
infrastructure. Our calibration layer is a contribution to that
defensive surface: an evidence producer that lives at the import-time
boundary and emits posteriors a downstream consumer can route on. The
calibrated probability is the primitive that ``operationalises trust''
in Pascual's terminology, and the metadata channel
(Section~\ref{sec:method-calibration}) is the answer the binary
registry probe cannot give to Seletskiy's question of
``who built it, what's in it, and whether it can be trusted.''

\paragraph{Calibration in machine learning.}
Guo et~al.~\cite{guo2017calibration} established the calibration of
modern neural networks as a first-order quality concern, demonstrating
that over-parameterised models exhibit systematic over-confidence and
introducing temperature scaling as the canonical post-hoc fix. They
also formalised the Expected Calibration Error (ECE) reliability
diagram methodology that we adopt in
Section~\ref{sec:eval-metrics}. Subsequent work has extended
calibration evaluation to structured outputs and beyond classification,
but the canonical primitives (binning, reliability diagrams, ECE)
trace to Guo et al. Our Beta calibration layer departs from
temperature scaling in three ways: (i)~the prior is explicit rather
than implicit in network weights, allowing audit and revision;
(ii)~the prior's epistemic status is classified into the 3-category
taxonomy and the classification is part of the artefact, not an
informal annotation; (iii)~the conjugate Beta-Binomial update means
the posterior evolves analytically under accumulated evidence,
whereas temperature scaling fixes the calibration at a single
held-out split.

\paragraph{Bayesian methods in software engineering.}
Furia, Feldt and Torkar~\cite{furia2021bayes} surveyed Bayesian Data
Analysis methodology in empirical software engineering, arguing for
explicit probabilistic reasoning over null-hypothesis significance
testing on per-study analyses. Their argument applies to studies of
software phenomena (effect sizes, treatment-comparison studies); our
work extends the same probabilistic discipline from study-level
analysis to detector-level confidence calibration, with the additional
requirement that prior \emph{provenance} (the 3-category taxonomy) be
part of the audit trail. The methodological lineage at our institution
continues from MIDAS~\cite{herbold2015midas,hillah2016sac,hillah2017stt},
where Bayesian belief networks orchestrated functional-test selection
on SOA pipelines under noisy outcomes; the transposition to the
present setting is direct, with LLM-emitted code snippets in the role
of test executions and registry-probe outcomes in the role of test
verdicts.

\section{Methodology}
\label{sec:methodology}

\paragraph{Overview.} The methodology has three components that
compose end-to-end. (1)~A \emph{rule-based detection layer}, derived
from a Code Quality Scanning (CQS) skill, parses Python source for
\texttt{import} statements and emits per-rule findings against PyPI.
The detection layer is the unit of action: each finding is a
\texttt{(rule, package, snippet)} triple. (2)~A
\emph{3-category epistemic taxonomy} classifies each detection rule's
prior by provenance --- empirical calibration, constructive
determinism, or traced engineering judgement --- and constrains the
prior's effective sample size accordingly. The taxonomy is the unit of
audit: a reviewer reading a flag knows what kind of claim the
underlying prior is making. (3)~A \emph{Bayesian calibration layer}
maps each rule to a Beta prior under the taxonomy, computes the
posterior under accumulated evidence by Beta-Binomial conjugacy, and
emits $\Pr(H \mid E)$ for downstream consumers. The calibration layer
is the unit of output: detection findings leave the system as
posterior probabilities, not as binary flags.

\begin{figure}[t]
\centering
\begin{tikzpicture}[
    font=\footnotesize,
    node distance=6mm and 9mm,
    box/.style={draw, rounded corners=2pt, align=center, inner sep=3pt,
                minimum height=8mm, fill=blue!4},
    chan/.style={draw, rounded corners=2pt, align=center, inner sep=3pt,
                 minimum height=8mm, fill=blue!10},
    cal/.style={draw, rounded corners=2pt, align=center, inner sep=3pt,
                minimum height=9mm, fill=blue!16},
    tax/.style={draw, rounded corners=2pt, align=center, inner sep=3pt,
                minimum height=9mm, fill=orange!12},
    io/.style={draw, rounded corners=2pt, align=center, inner sep=3pt,
               minimum height=8mm, fill=gray!8},
    arr/.style={-{Stealth[length=2mm]}, thick},
    layer/.style={draw, dotted, rounded corners, inner sep=5pt},
  ]
  \node[io] (snip) {LLM-emitted\\snippet};
  \node[box, right=of snip] (ast) {AST import\\extraction\\$-$ stdlib};
  \node[chan, right=of ast, yshift=7.5mm] (reg)
       {\textbf{registry channel}\\PyPI probe\\$\to$ 404? $\to$ neighbour};
  \node[chan, right=of ast, yshift=-7.5mm] (meta)
       {\textbf{metadata channel}\\PyPI JSON $\to$ recency,\\author, descriptor};
  \node[cal, right=16mm of reg, yshift=-7.5mm] (calib)
       {\textbf{Bayesian}\\\textbf{calibration}\\posterior by\\Beta-Binomial};
  \node[tax, above=11mm of calib] (taxn)
       {\textbf{epistemic}\\\textbf{taxonomy}\\cat~1/2/3 prior\\provenance $+$ ESS cap};
  \node[io, right=of calib] (out) {posterior\\$\Pr(H\mid E)$\\$\to$ CI gate};
  \draw[arr] (snip) -- (ast);
  \draw[arr] (ast.east) -- (reg.west);
  \draw[arr] (ast.east) -- (meta.west);
  \draw[arr] (reg.east) -- (calib.west);
  \draw[arr] (meta.east) -- (calib.west);
  \draw[arr] (taxn) -- (calib)
       node[midway, right=1pt, font=\scriptsize\itshape] {priors};
  \draw[arr] (calib) -- (out);
  \begin{scope}[on background layer]
    \node[layer, fit=(ast)(reg)(meta),
          label={[font=\scriptsize\itshape]above:detection layer~(1)}] {};
    \node[layer, fit=(taxn),
          label={[font=\scriptsize\itshape]above:epistemic taxonomy~(2)}] {};
    \node[layer, fit=(calib),
          label={[font=\scriptsize\itshape]below:calibration layer~(3)}] {};
  \end{scope}
\end{tikzpicture}
\caption{End-to-end workflow. Each LLM-emitted snippet is parsed to
top-level import names (standard library dropped); the strict-registry
and metadata channels emit per-rule findings; the Bayesian calibration
layer maps each rule to a Beta prior under the 3-category epistemic
taxonomy and emits a posterior $\Pr(H\mid E)$ that a downstream CI gate
thresholds. The ground-truth oracle (an alias-aware PyPI probe,
Section~\ref{sec:method-spec}) is intentionally \emph{separate} from
the detector predicates it scores.}
\label{fig:workflow}
\end{figure}
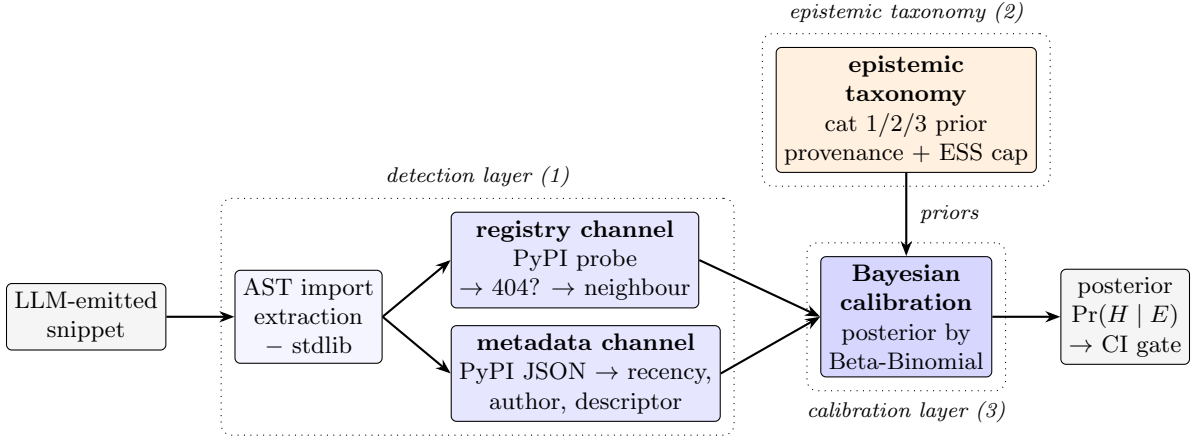

The detection layer is wrapped around a strict-registry channel
(probe PyPI, flag HTTP 404 results) and extended with a metadata
channel (read PyPI metadata, flag suspicious-but-registered packages).
The strict-registry channel addresses the canonical slopsquat regime
catalogued by Spracklen et~al.~\cite{spracklen2025packagehallu}; the
metadata channel addresses the
\emph{post-LLM-emission-attacker-registers} regime where the package
\emph{does} exist on PyPI but its metadata profile is atypical of a
legitimate dependency. Both channels share the calibration machinery.
Figure~\ref{fig:workflow} shows how the components compose, from an
LLM-emitted snippet to a per-finding posterior consumed by a CI gate.

\subsection{The Code Quality Scanning detector}
\label{sec:method-detector}

The detection layer parses Python source via AST, extracts top-level
import names, drops the standard library, and probes the PyPI JSON
registry for the remainder. Unresolved (HTTP 404) names are flagged.
For each flagged name, a single neighbour-search step queries
Levenshtein-distance-one candidates --- if any candidate \emph{is} on
PyPI, the flag is classified as a likely slopsquat (typo-squat-style
hallucination). The implementation reuses
ast-grep~\cite{astgrep-docs} for AST extraction, registry probing
through PyPI's public JSON endpoint, and an LRU cache for
repeated lookups. Spracklen et~al.~\cite{spracklen2025packagehallu}
report that approximately $97\%$ of LLM-emitted registry-miss imports
correspond to genuine hallucinations across the families they study;
this empirical observation provides contextual support (not parameter
calibration) for the registry-miss rule's prior discussed below.

\paragraph{Scope of the registry probe.} Two cases lie outside what a
registry-miss probe can decide, and we are explicit about both.
\emph{(i)~The hallucinated name is already registered as a slopsquat.}
If an adversary has pre-emptively registered the hallucinated name, the
PyPI probe returns HTTP $200$ and the strict-registry channel is
structurally blind to it. This is precisely the regime the metadata
channel (Section~\ref{sec:method-calibration}) is designed to surface,
and the empirical case in which it earns its keep
(Section~\ref{sec:results-discordance}): the package resolves, but its
metadata profile (recent registration, sparse author/summary, low
release count) is atypical of a legitimate dependency.
\emph{(ii)~A legitimate package whose own (transitive) imports have been
poisoned.} If \texttt{import requests} resolves to a genuine
distribution but an attacker has compromised one of \emph{its}
dependencies, no import-time probe of the top-level name can detect it:
our unit of analysis is the import name emitted by the LLM, not the
internal dependency closure of an existing distribution. We treat
transitive dependency compromise as out of scope and revisit it as a
construct-validity limitation in Section~\ref{sec:threats}; it is
complementary to lock-file-level provenance tooling
(\emph{e.g.}\ in-toto~\cite{torresarias2019intoto}, OSV-Scanner) rather
than something the calibration layer claims to address.

\subsection{Formal specification: oracle versus detector predicates}
\label{sec:method-spec}

To make the evaluation auditable we specify the detection rules as
typed predicates over an explicit state, separating the
\emph{ground-truth oracle} (which \emph{defines} the label) from the
\emph{detector predicates} (which \emph{emit} posteriors). Let the
scoring-time state be $\sigma = (R, I, \alpha, t)$, where $R$ is the
PyPI registry snapshot at corpus-build time, $I$ the multiset of
top-level import names extracted from a snippet, $\alpha$ the curated
import-name $\to$ distribution-name alias map
(Section~\ref{sec:results-adv}), and $t$ the snapshot timestamp. Write
$\textsc{resolve}_\alpha(p, R) \equiv [\,p \in R \,\lor\, \alpha(p) \in R\,]$
for the alias-aware resolution of a name $p$.

\emph{Oracle predicate (defines ground truth).}
$\textsc{halluc}(p) \equiv \lnot\,\textsc{resolve}_\alpha(p, R)$ ---
a name is labelled hallucinated iff neither it nor its alias resolves in
$R$ at time $t$.

\emph{Detector predicates (emit findings).} The strict-registry rules
are $\texttt{registry\_miss}(p) \equiv \lnot\,\textsc{resolve}_\alpha(p,R)$
and
\[
  \texttt{slopsquat\_neighbour}(p) \equiv
  \texttt{registry\_miss}(p) \land \exists q\, [\,\mathrm{lev}(p,q)=1
  \land \textsc{resolve}_\alpha(q,R)\,];
\]
the metadata rules are
predicates over the registered package's PyPI JSON
$M(p)$ (registration age $t - t_{\mathrm{reg}}(p)$, release count,
author and summary descriptors), defined in
Section~\ref{sec:method-calibration}.

This separation makes one fact precise. The strict-registry detector
predicate $\texttt{registry\_miss}$ is \emph{definitionally identical}
to the oracle predicate $\textsc{halluc}$: both reduce to
$\lnot\,\textsc{resolve}_\alpha(p,R)$ on the same snapshot. A binary
detector built from $\texttt{registry\_miss}$ alone therefore agrees
with ground truth by construction, which is why we treat such a
detector as a degenerate oracle upper bound rather than a competitor
(Section~\ref{sec:eval-baseline}). The metadata predicates are the only
rules in the pipeline that are \emph{not} a function of
$\textsc{resolve}_\alpha$, and are consequently the only ones that can
disagree with the registry oracle --- the formal reason the
contribution lives on the metadata channel. The specification style
follows the predicate-over-state discipline of Lamport's
TLA$^{+}$~\cite{lamport2002specifying} and the model-checking
literature~\cite{baierkatoen2008mc}, applied here only to fix the
oracle/detector boundary, not to perform an exhaustive state-space
exploration.

\subsection{The 3-category epistemic taxonomy}
\label{sec:method-taxonomy}

Every Beta prior in the detection pipeline is annotated with an
epistemic category that records its provenance. Throughout, the
\emph{effective sample size} (ESS) of a $\textsc{Beta}(\alpha,\beta)$
prior is $\alpha+\beta$: the number of pseudo-observations the prior is
worth, and therefore how much real adjudicated evidence is needed to
overrule it under the conjugate update of
Section~\ref{sec:method-calibration}. A larger ESS makes a prior
``stickier''; the taxonomy ties each category to an ESS bound so that
weaker provenance yields to data sooner.

\paragraph{Category 1 --- empirical calibration verified.} The prior
$\textsc{Beta}(\alpha,\beta)$ is derived by moment-matching or
quantile-matching from peer-reviewed statistics that measure the
modelled quantity directly. The source is cited \emph{as evidence
for the parameter values}. \emph{Constructive example.} If a
peer-reviewed study reported the precision of a particular detection
rule as $0.92$ over $N = 500$ adjudicated findings, the cat.~1 prior
would be $\textsc{Beta}(\alpha = 0.92 \cdot 500 = 460,
\beta = 0.08 \cdot 500 = 40)$ with $\textsc{ESS} = 500$ inherited
from the source. The cited paper is the source of the parameters,
and a future reviewer can audit the moment-matching directly.

\paragraph{Category 2 --- constructive deterministic argument.} The
prior derives from a formal property of the detection mechanism
itself rather than from an empirical measurement of the rule's
precision. \emph{Constructive example.} A registry probe that returns
HTTP $404$ from PyPI's JSON endpoint is, modulo network or
eventual-consistency issues, a hard statement about the absence of
the name from the registry: the precision of a registry-miss flag is
bounded by registry correctness, not by a heuristic. The cat.~2
prior $\textsc{Beta}(98, 2)$ with $\textsc{ESS} = 100$ encodes
``near-certain'' detection with a small residual that absorbs
transient mis-probes; the parameters trace to the constructive
argument, not to a peer-reviewed precision measurement of the rule.

\paragraph{Category 3 --- traced engineering judgement.} The prior
is a conservative baseline informed by industrial knowledge but
\emph{not} empirically calibrated. Contextual references (vendor
reports, related studies) are marked \emph{as context}, not as the
source of the parameter values. \emph{Constructive example.} Suppose
an engineer notes that ``recently-registered single-release PyPI
packages with empty author fields are suspicious in roughly $1$ in
$3$ cases I have seen during code review'' --- a domain-knowledge
hunch grounded in operational experience but never empirically
measured. The cat.~3 prior
$\textsc{Beta}(3, 7)$ with $\textsc{ESS} = 10$ encodes the hunch
(mean $0.30$) while explicitly limiting its weight against
accumulated evidence. Each cat.~3 prior carries a \emph{promotion
plan} --- a documented path to cat.~1 via instrumentation. The
threshold for automatic re-labelling
(cat.~3~$\rightarrow$~cat.~2) is set at $N \ge 20$ accumulated
observations, and a human-reviewed checkpoint at $N \ge 50$ tests
whether the original prior mean lies within the $95\%$ credible
interval of the current posterior --- triggering re-labelling to
cat.~1 if so, or recalibration otherwise. The thresholds derive from
the requirement that the posterior become data-driven before
re-labelling: at our cat.~3 ESS cap (below), $N = 20$ confirmations
make the posterior roughly two-thirds data-driven.

\paragraph{ESS bounds per category.} The categories also encode bounds
on the effective sample size of each prior, expressed as
$\alpha + \beta$. Cat.~1 priors inherit the $N$ of the source study.
Cat.~2 priors are capped at $\alpha + \beta \le 100$, reflecting the
strength of the constructive argument. Cat.~3 priors are capped at
$\alpha + \beta \le 10$, ensuring that traced engineering judgement
yields to roughly $20\text{--}50$ confirmed observations rather than
dominating the posterior indefinitely. The $10\times$ gap between
cat.~2 and cat.~3 caps is audit-readable: categories encode
quantified commitments on prior revisability under evidence, not
decorative labels.

\paragraph{On the cat.~2 cap of $100$.} We state plainly that the
specific value $\alpha+\beta\le 100$ is an engineering anchor, not a
measured quantity --- consistent with the taxonomy's own demand for
epistemic honesty. Its \emph{role} is derived: the cap must be large
enough that a near-deterministic registry probe is not casually
overturned by a handful of noisy adjudications, yet small enough that a
genuinely systematic divergence (say, a sustained run of
eventual-consistency mis-probes) can still move the posterior within an
operationally tractable number of observations. At $\textsc{ESS}=100$,
$20$ contradicting observations shift a $\textsc{Beta}(98,2)$ posterior
mean from $0.98$ to $\approx 0.82$ and $50$ shift it to $\approx 0.65$
--- revisable, but only under real evidence. The cap is therefore
\emph{revisable by construction}: the documented promotion path that
re-grounds a cat.~3 prior on instrumented data applies equally to
re-grounding this anchor on a measured PyPI 404-correctness rate, at
which point the cat.~2 prior is itself promoted to cat.~1 with the
source study's own $N$. We flag the value as an anchor rather than
launder it as a derived constant.

\subsection{Bayesian calibration layer}
\label{sec:method-calibration}

Each detection rule $r$ is paired with a prior
$\pi_r = \textsc{Beta}(\alpha_r, \beta_r, c_r, s_r)$ where
$(\alpha_r, \beta_r)$ are the Beta parameters, $c_r \in \{1,2,3\}$ is
the epistemic category, and $s_r$ is a citation key or design-decision
identifier recording the source.

\paragraph{Initial priors --- strict-registry channel.} For the
core registry-probe detector we distinguish three rules:

\begin{itemize}[noitemsep,topsep=0pt]
  \item \texttt{registry\_miss} --- the package name does not resolve on
        PyPI. Prior: $\textsc{Beta}(98, 2)$ (mean $0.98$,
        $\textsc{ESS}=100$), classified \textbf{cat.~2}. The
        constructive argument is that an HTTP $404$ from PyPI's JSON
        endpoint is, modulo transient infrastructure issues, a
        deterministic statement about the absence of the name from the
        registry --- the precision of a registry-miss flag is bounded
        by registry correctness, not by a heuristic. Spracklen
        et~al.~\cite{spracklen2025packagehallu} provide independent
        empirical context (prevalence of LLM-emitted hallucinations) but
        are cited \emph{as context for the threat model}, not as the
        source of the parameter values.
  \item \texttt{slopsquat\_neighbour} --- a registry-miss plus a
        Levenshtein-distance-one neighbour that does exist on PyPI.
        Prior: $\textsc{Beta}(99, 1)$ (mean $0.99$,
        $\textsc{ESS}=100$), classified \textbf{cat.~2}. The
        constructive argument extends \texttt{registry\_miss}: the
        neighbour search is itself deterministic, so a near-neighbour
        on PyPI co-existing with the missing name is a structural
        signal of typo-squat-style emission rather than a heuristic
        inference. A small residual uncertainty is preserved by the
        finite ESS rather than hard-coded to confidence $1.0$ as in
        earlier rule-based detectors.
  \item \texttt{unreachable} --- the registry probe failed (network
        error, rate limit, transient infrastructure). Prior:
        $\textsc{Beta}(1, 1)$ (uniform, mean $0.5$,
        $\textsc{ESS}=2$), classified \textbf{cat.~2} as a
        principled no-information prior. A downstream consumer can
        decide whether to treat the probe failure as risk or as no-data
        from the posterior alone.
\end{itemize}

\paragraph{Initial priors --- metadata channel.} The strict-registry
channel can only flag names that do not appear on PyPI. The realistic
post-LLM-emission attack registers a hallucinated name shortly after
seeing the LLM hallucinate it, so the package \emph{does} resolve on
PyPI but its metadata profile (recent registration, single release,
empty or placeholder author, single-word summary) is atypical of a
legitimate dependency. We extend the calibration layer with three
metadata-aware rules that complement the strict-registry channel:

\begin{itemize}[noitemsep,topsep=0pt]
  \item \texttt{metadata\_recent\_single} --- the package is recently
        registered (within a configurable window before corpus build,
        set to $180$ days in our runs) and has
        a single release. Prior: $\textsc{Beta}(4, 6)$ (mean $0.40$,
        $\textsc{ESS}=10$), classified \textbf{cat.~3}. The prior
        captures the engineering intuition that a recent single-release
        package is suspicious but not conclusively malicious; routine
        new packages share this profile.
  \item \texttt{metadata\_\allowbreak recent\_\allowbreak empty\_\allowbreak author} --- recent registration
        plus an empty or placeholder author field. Prior:
        $\textsc{Beta}(3, 7)$ (mean $0.30$, $\textsc{ESS}=10$),
        classified \textbf{cat.~3}.
  \item \texttt{metadata\_recent\_low\_descriptor} --- recent
        registration plus an empty or single-word summary. Prior:
        $\textsc{Beta}(2.5, 7.5)$ (mean $0.25$, $\textsc{ESS}=10$),
        classified \textbf{cat.~3}.
\end{itemize}

A package can match multiple metadata-suspicion rules. We compose them
by taking the maximum posterior mean across triggered rules rather than
multiplying probabilities; multiplication assumes rule-independence
that the metadata signals plainly violate (recency drives several of
them).\footnote{The principled object here is the marginal posterior
$\Pr(H\mid E_1,\dots,E_k)$ of a Bayesian belief
network~\cite{pearl1988probabilistic,koller2009pgm} whose evidence
nodes $E_i$ (the triggered rules) are dependent through a shared
recency parent. We do not learn that network; the maximum-mean rule is
a deliberately conservative approximation to its marginal that is exact
in the mutually-exclusive-evidence limit and otherwise under-states
joint suspicion, erring toward fewer false positives. Replacing it with
a fitted belief network is a natural refinement once enough adjudicated
co-occurrences exist to estimate the couplings.} The maximum-mean
composition is conservative on the false-positive side and admits the
rule-class ablation reported in
Section~\ref{sec:results-v3}.

\paragraph{Posterior update.} As labelled outcomes (true positive
versus false positive) accumulate, each rule's posterior is updated by
Beta-Binomial conjugacy:
\[
   \textsc{Beta}(\alpha_r + s, \beta_r + f),
\]
where $s$ is the count of true-positive confirmations and $f$ the
count of false positives. The posterior mean replaces the prior mean
in the emitted detection confidence; the posterior $\alpha$ and
$\beta$ are preserved so downstream consumers can recover variance,
credible intervals, or the full distribution if needed.

\paragraph{Worked example --- the \texttt{pyzenith} case.} The
package \texttt{pyzenith} (PyPI, registered $2025\text{-}12\text{-}18$)
is a real public Python project advertising itself as a
``Cross-Platform ML Optimization Framework with ONNX''. It happens
to carry an empty \texttt{info.author} field and was first published
within the $180$-day window before our corpus build, so the metadata
channel's \texttt{metadata\_\allowbreak recent\_\allowbreak empty\_\allowbreak author} rule fires when a
panel LLM emits \texttt{import pyzenith}. All six panel models
emit this import on the matched stress-test prompt, and the
metadata channel flags all six snippets
(Section~\ref{sec:results-discordance}).

\begin{itemize}[noitemsep,topsep=0pt]
  \item \textbf{Mahmud-shaped binary.} The package resolves
        on PyPI (HTTP $200$); no flag is emitted.
  \item \textbf{CQS calibrated, registry channel only (variant~2).}
        Same outcome: no flag.
  \item \textbf{CQS calibrated, registry + metadata channels
        (variant~3).} The \texttt{metadata\_\allowbreak recent\_\allowbreak empty\_\allowbreak author}
        rule fires. The prior is $\textsc{Beta}(3, 7)$, mean $0.30$:
        the emitted output is a flag with posterior confidence
        $0.30$ --- well below the strict-registry channel's
        $\sim 0.98$, and well below any sensibly-set CI-gate threshold
        $\tau = 0.5$. A consumer treating the flag as a
        \emph{surface-for-review} rather than an \emph{auto-block}
        signal would route \texttt{pyzenith} into manual triage; an
        \emph{auto-block} consumer with $\tau = 0.5$ would not block
        it. The point is precisely that the consumer chooses, with
        the calibrated probability as input.
  \item \textbf{Posterior update.} Suppose manual triage adjudicates
        \texttt{pyzenith} as legitimate ($f = 1$, a false-positive
        observation for the rule). The posterior is
        $\textsc{Beta}(3, 8)$, posterior mean
        $3 / (3 + 8) \approx 0.27$: the rule's confidence drops
        slightly toward ``less suspicious than the prior thought''.
        After $k$ false-positive adjudications and no true-positives,
        the posterior becomes $\textsc{Beta}(3, 7 + k)$, mean
        $3/(10 + k)$; at $k = 20$ the rule's posterior mean is
        $\approx 0.13$ and the prior-vs-data weighting has flipped
        to data-dominated, triggering the cat.~3~$\rightarrow$~cat.~2
        re-labelling described above.
\end{itemize}

The example illustrates the structural lever calibration adds.
The metadata channel surfaces \texttt{pyzenith} as worth a second
look without committing the binary baseline's silence (and without
committing a hard binary flag that the registered-but-legitimate
package would suffer from). Whether \texttt{pyzenith} is genuinely
suspicious is decided downstream; the calibrated layer makes the
question askable. We revisit \texttt{pyzenith} as the most prominent of
the metadata-channel discordant flags on the full $N = 1734$ corpus
(Section~\ref{sec:results-discordance}).

\paragraph{Implementation.} The calibration layer is implemented in
roughly $400$ lines of Python: a \texttt{PosteriorBuilder} mapping rule
names to priors, a free function \texttt{update\_posterior} for the
conjugate update, and a metadata-suspicion module that reads PyPI's
JSON response and routes packages to the metadata rules described
above. The detector consults the builder at scoring time and surfaces
the posterior mean as its emitted confidence; alpha, beta, category,
and source are attached as metadata for downstream auditing. No
external libraries are required beyond the standard
\texttt{scipy.stats.beta} for credible-interval reporting and
\texttt{requests} for registry probing.

\section{Empirical evaluation setup}
\label{sec:evaluation}

\subsection{Corpus --- dual sourcing}
\label{sec:eval-corpus}

The corpus is sourced from two complementary prompt sets, chosen to
expose two distinct properties of slopsquat detection. The combined
\emph{prompt} set contains $289$ prompts: $189$ stratified from
BigCodeBench (the realistic-use slice, addressing industry-relevance
concerns) and $100$ from a complementary stress-test prompt set
described below. Each prompt is run through the six-model panel of
Section~\ref{sec:eval-panel}, and the parseable generations form the
evaluation corpus of $N=1{,}734$ snippets ($1134$ realistic-use $+$
$600$ stress-test, i.e.\ $289 \times 6$).

\paragraph{Realistic-use prompts (BigCodeBench).} We use Python
snippets stratified from
BigCodeBench~\cite{bigcodebench2025iclr}, a peer-reviewed benchmark of
$1140$ library-heavy Python tasks designed to stress-test LLM code
generation across diverse function-call patterns. Sampling is
stratified across six broad library-use categories (numerical, ML,
HTTP, security, visualisation, other) at a fixed seed for
reproducibility. These prompts reflect realistic developer use:
``compute X using library Y'' --- where the library is mainstream
and the model is likely to use it correctly.

\paragraph{Stress-test prompts (niche-library elicitation).} The
BigCodeBench prompts alone yielded a base hallucination rate near
zero across our LLM panel (Section~\ref{sec:results}), which is
itself an important empirical finding of this paper: modern
code-specialised LLMs on realistic-use prompts do not produce
slopsquats at scale. To exercise the detection layer in the
tail-risk regime where slopsquat detection matters, we add a
second prompt set targeting niche or invented library names
(e.g.\ ``\textit{Implement clustering using a library called
pyzenith}'', or ``\textit{Use the deeplog package for log anomaly
detection}''). We frame this slice explicitly as a
\emph{stress-test corpus}: its purpose is not to characterise the
typical-use distribution in industry --- which the realistic-use
slice already addresses --- but to provoke the failure mode under
study and exercise the detector in the regime where it has work to
do. Real-world incidents of adversaries registering hallucinated
names~\cite{spracklen2025packagehallu} confirm this regime is not
merely synthetic: a CI gate must catch the rare hallucination
\emph{before it reaches production}, not the common-case success.

Each task is processed through the LLM panel below; each generated
snippet is parsed for imports, top-level external names are extracted,
the standard library is filtered out, and each remaining name is
probed against PyPI's JSON API. Names returning HTTP $404$ are
labelled \emph{hallucinated}; names returning HTTP $200$ are
labelled \emph{real}; HTTP errors are treated conservatively (assume
real).

\subsection{LLM panel}
\label{sec:eval-panel}

\begin{table}[h!]
\centering
\small
\begin{tabular}{llll}
\toprule
Developer & Model & Mode & Role \\
\midrule
Anthropic & claude-sonnet-4.6   & cloud (OpenRouter)  & Closed-source cloud comparator \\
Mistral   & mistral-large-2411  & cloud (OpenRouter)  & Cloud general-purpose, European \\
DeepSeek  & deepseek-v4-pro     & cloud (direct API)  & Cloud reasoning anchor \\
DeepSeek  & deepseek-r1         & cloud (OpenRouter)  & Within-family reasoning pair \\
Mistral   & codestral:22b       & local (ollama)      & Code-specialised, European \\
Meta      & codellama:7b        & local (ollama)      & Code-specialised, small \\
\bottomrule
\end{tabular}
\caption{LLM panel: $6$ models across $4$ developers spanning four
cloud models and two local open-weight code models. DeepSeek
contributes a within-family pair (general-purpose \texttt{v4-pro} plus
reasoning \texttt{r1}) and Mistral contributes a cloud / local pair
(\texttt{mistral-large} plus the code-specialised \texttt{codestral}).
The cloud / local split exposes whether the calibration generalises
across deployment mode; the developer span exposes whether it
generalises across organisations, geographies, and training-data
distributions.}
\label{tab:panel}
\end{table}

The panel is designed to expose two orthogonal sources of variance.
\emph{Cross-developer diversity} (rows of Table~\ref{tab:panel}) tests
whether the calibration generalises across organisations,
geographies, and training-data distributions, with one closed-source
cloud model (Claude-Sonnet-4.6) included to address the
realistic enterprise distribution of models in production use.
\emph{Cloud vs.\ local diversity} tests whether the calibration
generalises across deployment mode (API-billed cloud models versus
locally-hosted open-weight code models). We adopt the inference
parameters $\text{temperature}=0.2$, $\text{top\_p}=0.9$,
$\text{max\_tokens}=2048$, and pin model versions and access dates
(Section~\ref{sec:reproducibility}). The two local open-weight models
are reproducible at the byte level from their published GGUF artefacts;
the four cloud-routed models are reproducible from any account with the
equivalent routes (Section~\ref{sec:reproducibility} discusses the
cloud-decoding reproducibility caveat).

\subsection{Baseline}
\label{sec:eval-baseline}

\paragraph{Why a re-implementation.} Mahmud
et~al.~\cite{mahmud2025trustcalibrated} do not publish a public
reference implementation of their pipeline; the IEEE Xplore record
carries no code-availability statement and we found no associated
public repository at the authors' institutions (Tuskegee University
and U.S.\ Army DEVCOM Armaments Center) at the time of writing. A
fidelity-quantified replication of their full multi-stage pipeline
would therefore require either authors-side code release or a careful
inference-from-figures replication that is beyond the scope of this
paper. We adopt the next-best comparator: a single-stage binary
detector that implements the strict-registry leg of the Mahmud
pipeline as described in their text (parse imports, drop standard
library, probe registry, flag unresolved). This baseline is the
natural single-stage detector their architecture would degrade to in
the absence of the trust-propagation stages, and the natural
lower-bound any practical slopsquat detector would reproduce. We
frame the comparison throughout as \emph{calibrated CQS} versus
\emph{Mahmud-shaped binary} rather than \emph{ours versus
Mahmud~2025} to avoid overclaiming fidelity. The fidelity gap is
bounded above by the trust-propagation stages we omit, which act on
already-flagged findings rather than on the strict-registry channel
itself, and we surface it explicitly as a threat in
Section~\ref{sec:threats}.

\paragraph{What the baseline emits.} The baseline emits a binary
$\{\text{flagged}, \text{not flagged}\}$ decision per import, with
no confidence score by design. This is precisely the property our
calibration layer adds and the empirical lever the comparison
exercises.

\subsection{Metrics}
\label{sec:eval-metrics}

\paragraph{Precision, Recall, $F_1$.} For each detector variant we
threshold the emitted confidence at $\tau = 0.5$ to recover a binary
flag, then compute standard precision (true positives over total
flags), recall (true positives over total ground-truth hallucinations),
and their harmonic mean $F_1$. The choice $\tau = 0.5$ is the natural
Bayes threshold for a $0/1$-loss decision; consumers with asymmetric
costs (false-positive friction vs.\ missed-hallucination risk) can
re-threshold the posterior at any $\tau \in [0, 1]$.
The example consumer who wants to surface \texttt{pyzenith}-like
metadata flags for review but auto-block only strict-registry flags
would set $\tau_{\text{review}} = 0.20$ and
$\tau_{\text{block}} = 0.90$.

\paragraph{McNemar paired test.} The McNemar
test~\cite{mcnemar1947} is the standard non-parametric test for whether
two detectors disagree \emph{systematically} on paired data --- here,
paired by snippet. The contingency table records $a$ (both detectors
agree positive), $d$ (both agree negative), $b$ (calibrated flags,
binary does not), and $c$ (binary flags, calibrated does not). Only the
\emph{discordant} pairs $(b,c)$ carry information about asymmetry: under
the null of no systematic difference, each discordant pair is equally
likely to fall in $b$ or $c$, so the test reduces to a two-sided
binomial test on $\min(b,c)$ out of $b+c$ trials at $p=\tfrac12$. We use
the exact binomial form, which is appropriate for the small discordant
counts here ($b+c \le 49$); for the modern guidance on McNemar variants
(exact-conditional vs.\ mid-$p$ vs.\ asymptotic, and when each is
preferable) we follow Fagerland
et~al.~\cite{fagerland2013mcnemar}. We report $b$, $c$, the two-sided
$p$-value, and the qualitative interpretation; the
\texttt{pyzenith}-style metadata-only flags contribute exactly to
$b$ (calibrated flags, binary does not).

\paragraph{Expected Calibration Error (ECE).} Following
Guo et~al.~\cite{guo2017calibration}, ECE is the weighted mean
absolute gap between predicted confidence and empirical accuracy,
binned over the confidence range:
\[
   \mathrm{ECE} = \sum_{m=1}^{M} \frac{|B_m|}{N}
       \left| \mathrm{acc}(B_m) - \mathrm{conf}(B_m) \right|,
\]
where $B_m$ is the $m$-th confidence bin, $|B_m| / N$ is its
relative weight, $\mathrm{acc}(B_m)$ is the empirical accuracy of
predictions in the bin, and $\mathrm{conf}(B_m)$ is the average
predicted confidence in the bin. We use $M = 10$ equal-width bins
over $[0, 1]$. \emph{Two views.} The \textbf{full-corpus ECE} weights
every snippet equally, including the $\sim 80\%$ where the detector
correctly emits no flag and zero confidence; on a clean-dominated
corpus this view is degenerate (the trivial all-zero predictor
already achieves a very small full-corpus ECE) and we report it for
transparency only. The \textbf{flagged-subset ECE} restricts the
computation to the snippets the detector flagged, the conventional
calibration view for a detector. We report both.

\paragraph{Brier score.} Because the flagged-subset ECE conditions on
positive predictions, it is silent on the cost of \emph{missed}
hallucinations and on the over-confidence of the un-flagged mass. We
therefore also report the Brier score~\cite{brier1950score} --- the
mean squared error between the predicted probability of the positive
class (hallucinated) and the binary outcome --- over the \emph{full}
corpus, with un-flagged snippets contributing a predicted probability
of $0$:
\[
   \mathrm{Brier} = \frac{1}{N}\sum_{i=1}^{N}\bigl(\hat{p}_i - y_i\bigr)^2 ,
\]
where $\hat{p}_i$ is the detector's posterior $\Pr(H\mid E)$ for
snippet $i$ and $y_i \in \{0,1\}$ its ground-truth label. Unlike the
flagged-subset ECE, the Brier score is a strictly proper scoring rule
evaluated over every snippet, so it penalises both false flags and
silent misses; it complements rather than replaces the conventional
calibration view.

\paragraph{Reliability diagrams and per-model ECE.} On the flagged
subset the reliability diagram plots, per confidence bin, the empirical
\emph{precision} (the fraction of flags in the bin that are true
hallucinations under binary ground truth) against the mean predicted
confidence; a perfectly-calibrated detector traces the diagonal
(Figure~\ref{fig:reliability}). We additionally check calibration
robustness across the LLM panel through the per-model precision
break-out (Table~\ref{tab:results-per-model}): a calibration that held
in aggregate but drifted on an individual model would surface as a
per-model precision outlier, which is exactly how the
\texttt{claude-sonnet-4.6} registry-channel precision is read in
Section~\ref{sec:results-v3}. Full per-model ECE values are retained in
the reproducibility material.

\paragraph{Power analysis.} The corpus size was fixed \emph{a priori}
by a standard McNemar power calculation. Setting two-sided
$\alpha = 0.05$, target power $0.80$, and expected
detector-disagreement rate $p_{\text{disagree}} = 0.10$ (the rate
of discordant pairs as a fraction of paired snippets), a design target
of $\approx 300$ paired snippets detects a
relative asymmetry of approximately $50\%$ on the discordant-pair
counts --- equivalent to an absolute $b/c$ gap of about $5\%$ ---
which is the smallest practically meaningful effect. The realised
corpus of $N=1734$ paired snippets (six models $\times$ $289$ prompts)
was sized to clear that target with margin. Post hoc, however, all
three detector comparisons in Table~\ref{tab:results-mcnemar} turn out
to be one-directional by construction (Section~\ref{sec:results-adv}):
the realised discordance rates ($39$, $49$ and $10$ out of $1734$, i.e.
$0.6$--$2.8\%$) all fall well below the $10\%$ assumption, and corpus
size does not confer McNemar power when the relevant sample is the
discordant count and the reverse cell is structurally empty. We
therefore read these counts descriptively
(Section~\ref{sec:threats}) rather than as powered hypothesis tests.
The earlier $N=91$ pilot surfaced $b=3, c=0$ on the
variant-2-vs-variant-3 pairs; the full corpus surfaces $b=10, c=0$.

\section{Results}
\label{sec:results}

We report results from the full $N=1734$ corpus: $1134$ BigCodeBench
(realistic-use) snippets and $600$ stress-test snippets, with
$n_{\mathrm{hallucinated}}=244$ under the alias-aware ground truth
(Section~\ref{sec:eval-corpus}), across the
six-model panel of Table~\ref{tab:panel}
(\texttt{claude-sonnet-4.6},
\texttt{mistral-large-2411},
\texttt{deepseek-v4-pro},
\texttt{deepseek-r1},
\texttt{codestral:22b},
\texttt{codellama:7b}). Each model contributes $189 + 100 = 289$
parseable snippets to the merged corpus.

The section is organised around a structural result and a case
study that explains it. The \emph{structural} result is that the
metadata-augmented detector (variant~3) raises exactly ten flag
decisions the registry-only calibrated detector (variant~2) does not,
and none in the other direction ($b=10$, $c=0$). This direction is not
a chance asymmetry: variant~3's rule set contains variant~2's, and
rules compose by maximum posterior mean
(Section~\ref{sec:method-calibration}), so variant~3's posterior
dominates variant~2's on every snippet and the reverse-discordance cell
is empty \emph{by construction}. We therefore report the McNemar
contingency descriptively and defer its inferential interpretation to
Section~\ref{sec:threats}. The \emph{case study}
(Section~\ref{sec:results-discordance})
traces those ten metadata-only flags to two registered-but-suspicious
PyPI packages; it illustrates the mechanism by which calibration changes
outcomes, and we are careful to present it as an existence proof rather
than a population-level prevalence claim. The binary $F_1$ comparison ranks the
metadata-aware detector below the strict-registry detectors; the threat
model (Section~\ref{sec:intro}) inverts that ranking. The remainder of
this section quantifies the disagreement and the calibration evidence
behind it.

\subsection{Three detectors disagree at \texorpdfstring{$N=1734$}{N=1734}}
\label{sec:results-adv}

We evaluate three detector variants on the merged corpus: a
Mahmud-shaped binary detector (strict registry probe only,
variant~1); a Bayesian-calibrated detector restricted to the
strict-registry channel (variant~2); and a Bayesian-calibrated
detector that adds the metadata channel (variant~3,
Section~\ref{sec:method-calibration}). The ground-truth label is
the binary outcome of an alias-aware PyPI registry probe at
corpus-build time (2026-05-23): an import name is labelled
non-hallucinated if either the raw name or its canonical
distribution-name alias (a curated $15$-entry whitelist covering
\texttt{dateutil}~$\rightarrow$~\texttt{python-dateutil},
\texttt{cv2}~$\rightarrow$~\texttt{opencv-python},
\texttt{PIL}~$\rightarrow$~\texttt{Pillow}, \emph{etc.}) returns
HTTP $200$. The same alias-aware probe is consumed by the
strict-registry channel of variants~1 and~2; ground truth and
detector therefore see the same registry semantics on this corpus,
and the headline detector comparison is not distorted by
well-known import-name vs.\ distribution-name pairs (cf.\
\texttt{IMPORT\_TO\_DISTRIBUTION} in the reproducibility material,
Section~\ref{sec:reproducibility}).

\begin{table}[h!]
\centering
\small
\resizebox{\textwidth}{!}{%
\begin{tabular}{lccccc}
\toprule
Detector & Precision & Recall & $F_1$ & Flagged ECE & Brier \\
\midrule
\emph{Mahmud-shaped binary (v1, oracle bound)} & \emph{1.000} & \emph{1.000} & \emph{1.000} & \emph{n/a} & \emph{n/a} \\
CQS-calibrated, registry channel only (variant~2)         & $0.956$ & $0.881$ & $0.917$ & $0.016$ & $0.022$ \\
CQS-calibrated, registry $+$ metadata channel (variant~3) & $0.915$ & $0.881$ & $0.898$ & $0.029$ & $0.023$ \\
\bottomrule
\end{tabular}}
\caption{Detection metrics on the merged $N=1734$ corpus
($1134$~BCB $+$ $600$~stress-test) across the six-model panel of
Table~\ref{tab:panel}, computed against the alias-aware ground
truth (Section~\ref{sec:eval-corpus}). \emph{The variant~1 row is
italicised because it is not a competitor}: a binary detector built
from the registry-miss predicate is definitionally identical to the
ground-truth oracle (Section~\ref{sec:method-spec}), so its
$F_1 = 1.000$ is a degenerate upper bound, not evidence of detection
quality. The substantive comparison is variant~2 vs.\ variant~3.
Brier is the strictly proper full-corpus score
(Section~\ref{sec:eval-metrics}). Variant~3 adds exactly $10$ flags
relative to variant~2, all classified as false positives under the
binary ground truth; Section~\ref{sec:results-discordance} reports
their identities and revisits the ground-truth definition.}
\label{tab:results-headline}
\end{table}

\paragraph{McNemar contingency on the binary detection decisions.}
Table~\ref{tab:results-mcnemar} reports the paired contingency for the
three detector pairs. In every pair one detector's flag set is nested
in the other's --- the calibrated variants share the registry channel
and differ only by added rules
(Section~\ref{sec:method-calibration}), and the Mahmud-shaped binary is
the registry oracle the calibrated variants approximate --- so in each
comparison all discordance falls in a single direction (the additional
flags of the richer or oracle-identical detector) and the opposite cell
is empty by construction rather than by chance. We therefore read these
counts \emph{descriptively}, as the number of flag decisions on which
the richer (or oracle-identical) detector differs, and treat the
accompanying exact-binomial $p$-values as nominal; their inferential
interpretation is qualified in Section~\ref{sec:threats}.

\begin{table}[h!]
\centering
\small
\begin{tabular}{lcccc}
\toprule
Comparison & $a$ & $b$ & $c$ & nominal $p$ \\
\midrule
v1 vs.\ v2 (Mahmud vs.\ registry-calibrated)   & $1695$ & $39$ & $0$ & $3.6 \times 10^{-12}$ \\
v1 vs.\ v3 (Mahmud vs.\ metadata-augmented)    & $1685$ & $49$ & $0$ & $3.6 \times 10^{-15}$ \\
\textbf{v2 vs.\ v3 (registry vs.\ metadata)}   & $1685$ & $10$ & $0$ & $\mathbf{0.0020}$ \\
\bottomrule
\end{tabular}
\caption{Three-way McNemar paired test on detector decisions.
The first two rows measure how far the calibrated detectors fall
\emph{short of} the degenerate oracle bound (variant~1) --- the flags
the oracle-identical binary raises that the calibrated detectors, by
design, attach a sub-threshold posterior to --- and are reported for
completeness, not as a head-to-head against a competitor. \textbf{The
third row is the result of interest}: the discordance the metadata
channel was designed to elicit (Section~\ref{sec:method-calibration}),
which the $N=91$ pilot did not surface ($b=3, c=0$). At
$N=1734$ the metadata channel produces $b=10$ flag decisions the
registry-only calibrated detector does not, and $c=0$ in the other
direction. Because the rule sets are nested and composed monotonically,
$c=0$ holds by construction; we report the exact-binomial $p$-values as
nominal figures and read the contingency descriptively
(Section~\ref{sec:threats}).}
\label{tab:results-mcnemar}
\end{table}

\subsection{The 10 discordant flags: pyzenith and chartcraft}
\label{sec:results-discordance}

The $10$ snippets that variant~3 flags and variant~2 does not
trace to exactly two PyPI packages. Six flags name
\texttt{pyzenith}; four flags name \texttt{chartcraft}. Each
package is emitted by multiple panel models in the stress-test
corpus, in response to prompts that explicitly name the library
(e.g.\ ``\textit{Implement Python code using a niche library called
pyzenith for clustering}'').

\begin{table}[h!]
\centering
\small
\begin{tabular}{lllc}
\toprule
Package & Registered on PyPI & Releases / author & Models that emit it \\
\midrule
\texttt{pyzenith}   & $2025\text{-}12\text{-}18$ & $18$\,/\,empty & all~$6$ panel models \\
\texttt{chartcraft} & $2026\text{-}03\text{-}19$ & $\phantom{0}2$\,/\,empty & $4$ of $6$ panel models \\
\bottomrule
\end{tabular}
\caption{The two PyPI packages responsible for the $10$
discordant flags. Both packages were registered on PyPI within the
$180$-day window before the corpus build
($2026\text{-}05\text{-}23$), well after the cut-off of the
training data on which the panel models were trained, and both
present the metadata profile that the calibrated metadata channel
flags as suspicious: recent registration and empty
\texttt{info.author}. Binary ground truth labels
both packages as ``not hallucinated'' because PyPI resolves the
name. The metadata-aware detector flags both as suspicious at
posterior confidence $\approx 0.30$ (the prior mean of the
\texttt{metadata\_\allowbreak recent\_\allowbreak empty\_\allowbreak author} rule).}
\label{tab:results-supplychain}
\end{table}

This case study illustrates the mechanism behind the statistical
headline; we present it as an existence proof of the failure mode, not
as a prevalence estimate ($n=10$ flags from two packages cannot carry a
population claim). The strict-registry
channel and binary ground truth share a single oracle (``does PyPI
resolve this name today''); they are therefore mechanically blind
to the attack pattern in which a hallucinated name is registered
on PyPI by an attacker shortly after a model learns to emit it. The
metadata channel is the only signal in our pipeline that
distinguishes a fresh-and-thin registration from a legitimate
dependency; it pays an $F_1$ cost relative to the binary metric
($\Delta F_1 = -0.019$ on the merged corpus, $\Delta P = -0.041$,
$\Delta R = 0$) and earns it back in
threat-model-aligned recall over the post-emission-registration
scenario --- exactly the regime catalogued by Spracklen
et~al.~\cite{spracklen2025packagehallu} as the highest-consequence
slopsquat tail.

We deliberately do not adjudicate whether \texttt{pyzenith} and
\texttt{chartcraft} are themselves malicious. The PyPI summary
strings are plausible (``Cross-Platform ML Optimization Framework
with ONNX Interpreter''; ``Python-powered dashboards that rival
Power BI \& Tableau'') and may correspond to legitimate-but-young
projects. The point is precisely that the binary registry probe
cannot tell, and a downstream consumer should not be required to
trust the registry probe's silence in the face of an
LLM-recommended import whose metadata footprint matches a
slopsquat profile. The calibrated posterior at $\approx 0.30$ is
designed to be a \emph{surface-for-review} signal, not an
\emph{auto-block} signal, and a CI gate at $\tau = 0.5$ would still
let both packages through.

\subsection{BigCodeBench (realistic-use): the tail-risk regime confirmed}
\label{sec:results-bcb}

On the $1134$-snippet BigCodeBench stratum, the per-model
hallucination rate is \emph{exactly zero} under our alias-aware
ground truth (Section~\ref{sec:eval-corpus}): no model emits an
import that fails to resolve on PyPI either directly or under
its canonical distribution-name alias. The two non-alias-aware
hallucinations that surface on this stratum
(\texttt{dateutil}~$\times$~$12$ and \texttt{cv2}~$\times$~$12$,
identical counts across every panel model) are entirely
import-name versus distribution-name mismatches:
\texttt{dateutil} resolves to the \texttt{python-dateutil}
distribution and \texttt{cv2} resolves to the
\texttt{opencv-python} distribution, both legitimate and widely
deployed. Our ground-truth labeller carries a curated
import-name $\rightarrow$ distribution-name alias map for the
fifteen most common Python projects where these two names differ;
the alias map is also consumed by the strict-registry channel
of the calibrated detector so that variant~1 and variant~2 see
the same registry semantics as ground truth.

The empirical headline is therefore strengthened relative to the
pilot ($N=91$ found one isolated codellama hallucination on the
BCB stratum at $5.0\%$ rate): \emph{modern code-specialised LLMs
on realistic-use prompts essentially do not produce strict
hallucinations once known import-name aliases are accounted for}.
Detection therefore matters precisely where the strict-registry
channel is structurally blind --- the metadata channel.

\subsection{Per-model robustness}
\label{sec:results-v3}

The discordance is panel-wide. \texttt{pyzenith} fires on
\emph{every} model in the panel: each of the six models produces
exactly one snippet in which \texttt{pyzenith} is the recommended
clustering library, and each such snippet is flagged by variant~3
and not by variant~1 or~2. \texttt{chartcraft} fires on four of the
six models. The metadata channel's discordance signal is therefore
not driven by a single outlier model but by a panel-wide tendency
to lean on the prompt's mention of the niche library name. The
strict-registry channel cannot see this because the squatter has
already registered the name.

The headline per-model $F_1$ table follows
(Table~\ref{tab:results-per-model}). The \texttt{claude-sonnet-4.6}
split is the registry-channel precision outlier: its small
hallucinated subset on the stress-test stratum (24 packages out of
$289$, the lowest in the panel) combines with $9$ false-positive
flag decisions to yield a registry-channel precision of
$0.73$. The other five models all sit at registry-channel
precision $\geq 0.97$.

\begin{table}[h!]
\centering
\small
\begin{tabular}{lccccc}
\toprule
Model & $\widehat{r}_{\mathrm{adv}}$ & v2~$F_1$ & v2~$P$ & v3~$F_1$ & v3~$P$ \\
\midrule
\texttt{claude-sonnet-4.6}     & $0.083$ & $0.82$ & $0.73$ & $0.80$ & $0.71$ \\
\texttt{mistral-large-2411}    & $0.159$ & $0.96$ & $1.00$ & $0.93$ & $0.96$ \\
\texttt{deepseek-v4-pro}       & $0.176$ & $0.92$ & $1.00$ & $0.90$ & $0.96$ \\
\texttt{deepseek-r1}           & $0.125$ & $0.96$ & $1.00$ & $0.94$ & $0.97$ \\
\texttt{codestral:22b}         & $0.145$ & $0.90$ & $0.97$ & $0.88$ & $0.92$ \\
\texttt{codellama:7b}          & $0.156$ & $0.93$ & $0.98$ & $0.91$ & $0.93$ \\
\bottomrule
\end{tabular}
\caption{Per-model detection metrics on the alias-aware ground
truth. $\widehat{r}_{\mathrm{adv}}$ is the merged $289$-snippet
hallucination rate; v2 is the registry-only calibrated detector;
v3 is the metadata-augmented detector. The metadata channel costs
$2\text{--}5$ precision points per model and never reduces recall.}
\label{tab:results-per-model}
\end{table}

\subsection{Calibration quality}
\label{sec:results-calibration}

The conventional reliability metric for a detector is computed on
its positive predictions (\emph{flagged-subset ECE}). The
registry-channel variant emits posterior confidence
$\approx 0.97\text{--}0.98$ on each of its $225$ strict-registry
flags against an empirical precision of $0.956$, yielding
$\mathrm{ECE} \approx 0.016$ --- a factor of six inside the
$0.10$ threshold adopted from Guo
et~al.~\cite{guo2017calibration}. The metadata-augmented variant
emits a broader range of posterior confidences (cat.~3 priors at
$0.25\text{--}0.40$ prior mean) over its $235$ flags, against an
empirical precision of $0.915$, yielding
$\mathrm{ECE} \approx 0.029$. The metadata channel pays an ECE
penalty of $\approx 0.013$ for broadening the calibration curve
across the confidence range; both variants remain well inside the
calibration target.

\begin{figure}[h!]
\centering
\includegraphics[width=0.7\columnwidth]{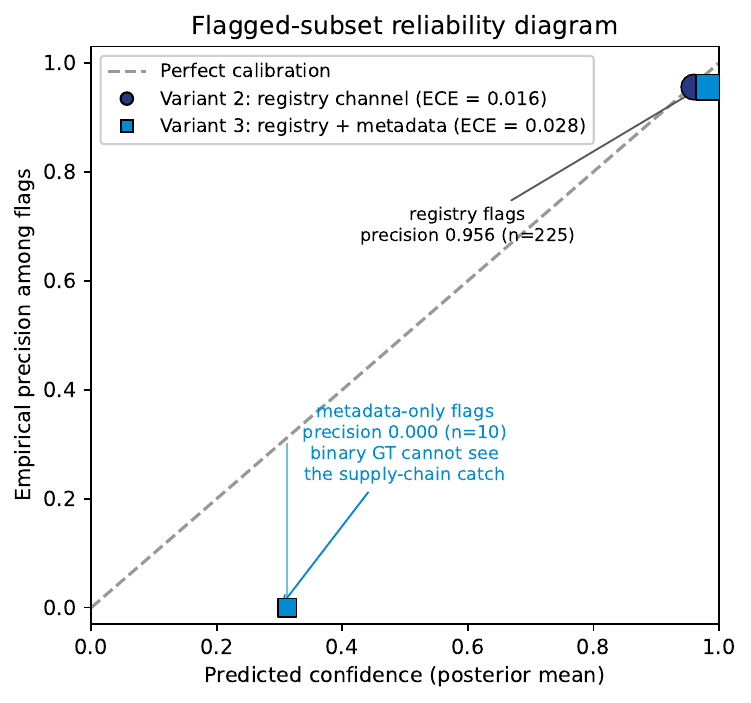}
\caption{Flagged-subset reliability diagram for variants~2
(registry channel only) and~3 (registry $+$ metadata channels) on
the $N=1734$ corpus under the alias-aware ground truth. Each marker is
an occupied confidence bin, plotted at (mean predicted confidence,
empirical precision among flags) with a thin line showing the gap to
the diagonal; marker area grows with the bin's flag count. Both
variants place their $225$ registry flags at confidence
$\approx 0.97$ and empirical precision $0.956$, on the diagonal
(well-calibrated). Variant~3 adds a single low-confidence bin at
posterior $0.30$ holding the ten metadata-only flags; under
\emph{binary} ground truth their empirical precision is $0.0$ (the
registry oracle cannot label the supply-chain catch as a positive), and
that gap is the entire source of variant~3's higher flagged-subset ECE
($0.029$ vs.\ $0.016$). Both ECEs remain within the $0.10$ target.}
\label{fig:reliability}
\end{figure}

The full-corpus Brier score (Section~\ref{sec:eval-metrics}) tells a
consistent story from the complementary direction: $0.022$ for the
registry channel and $0.023$ once the metadata channel is added. The
two scores are nearly identical because the metadata channel adds only
ten low-confidence ($0.30$) predictions over $1734$ snippets; the
strictly proper score confirms that broadening the calibration curve
does not degrade aggregate probabilistic accuracy. Note that the
flagged-subset ECE measures the metadata flags against \emph{binary}
ground truth, under which all ten are false positives by definition ---
so the metadata bin sits at empirical precision $0.0$ in
Figure~\ref{fig:reliability}. That gap is exactly the calibration
``penalty'' the channel pays for surfacing a catch the binary oracle
cannot label as a positive, not a defect in the posterior.

\subsection{Corpus health (V1)}
\label{sec:results-v1}

The per-model corpus-validity check (Section~\ref{sec:eval-metrics})
reports parse rates $\geq 91.7\%$ and uniqueness rates $\geq 97.6\%$
across the six-model panel. The mean number of external imports per
snippet ranges from $2.48$ (\texttt{codellama:7b}) to $2.98$
(\texttt{claude-sonnet-4.6}). The only flagged corpus property is
\texttt{halluc\_rate\_out\_of\_band} on the BigCodeBench-only
stratum, which is precisely the realistic-use tail-risk finding of
Section~\ref{sec:results-bcb} rather than a model defect.

\section{Discussion}
\label{sec:discussion}

\paragraph{Slopsquat detection is a tail-risk problem.} The
near-zero hallucination rate on BigCodeBench
(Section~\ref{sec:results-bcb}) is itself the most consequential
finding of this work. Modern code-specialised LLMs on realistic-use
prompts essentially do not produce slopsquats; deployed naively, a
slopsquat detector would emit no flags on the vast majority of
LLM-assisted commits. Yet the cost of \emph{one} unflagged slopsquat
slipping into a build is large (supply-chain compromise of every
downstream consumer). Detection therefore matters precisely in the
tail-risk regime where the base rate is low but the consequence is
catastrophic. A binary detector is poorly matched to this regime: it
either reports too many flags (high friction at the boundary) or too
few (missed compromise). A \emph{calibrated} detector is the natural
primitive --- every flag carries a probability that policy can route
on.

\paragraph{When calibration helps.} Calibrated confidence is most
valuable at the CI-gate boundary: instead of a single global
threshold, a downstream consumer (a pre-commit hook, a code-review
agent, or a security policy engine) can map confidence into a
policy decision specific to the context (developer machine versus
CI versus production deploy). The calibrated layer also composes
naturally with other risk signals: a Beta posterior can be combined
with other probabilistic indicators via standard probabilistic
operators rather than via ad-hoc voting rules.

\paragraph{When calibration matters less.} For headline detection
metrics on a clean corpus where the strict-registry channel suffices
(realistic-use BigCodeBench), the calibrated layer adds no
detections relative to the binary baseline. Its contribution there is
restricted to the posterior confidence on the (rare) flag and the
provenance metadata. The metadata channel earns its keep precisely
in the tail-risk regime where post-LLM-emission attacker registration
makes the strict-registry channel insufficient.

\paragraph{Robustness of the cat.~2 / cat.~3 split.} The
classification of strict-registry rules as cat.~2 rests on the
constructive determinism of registry probes. Two transient
phenomena complicate this: PyPI eventual-consistency windows
(within which a freshly-registered name may not yet resolve in
every region) and rate-limited probes (which surface as
\texttt{unreachable} rather than \texttt{registry\_miss}). Both are
handled by the existing \texttt{unreachable} rule, which is
deliberately uniform; we report the breakdown of flag categories
in the reproducibility material so that consumers can apply their
own posterior over transient-versus-genuine miss distinctions.

\section{Threats to validity}
\label{sec:threats}

\paragraph{Statistical conclusion validity.} Three features of the
detector design limit the inferential weight of the McNemar
contingencies in Section~\ref{sec:results-adv}, and we accordingly read
the metadata-channel discordance as an existence proof rather than a
powered test. \emph{First}, the comparison is one-directional by
construction: each richer detector's rule set contains the
comparator's and rules compose by maximum posterior mean
(Section~\ref{sec:method-calibration}), so the reverse-discordance cell
is pinned to zero and McNemar's symmetric null cannot obtain --- the
nominal $p$-values restate that the added channel produced flags, not
that a surprising asymmetry was observed. \emph{Second}, the discordant
flags are not independent matched pairs: the ten
variant-2-vs-variant-3 discordances trace to only two packages
(\texttt{pyzenith}, \texttt{chartcraft}), each emitted by several panel
models, so the effective number of independent discordant units is
closer to two than to ten. McNemar's test assumes pair-to-pair
independence; repeated measurements on the same unit violate it and
inflate nominal significance, and a cluster-aware variant is required
to recover a valid
test~\cite{eliasziw1991mcnemar,durkalski2003clustered}. \emph{Third},
the flag count underlying the contingency uses a rule-fired criterion
rather than the $\tau=0.5$ threshold used for the precision/recall/$F_1$
metrics (Section~\ref{sec:eval-metrics}); at $\tau=0.5$ the metadata
flags, whose posterior is $\approx 0.30$, are negative decisions (a CI
gate at $\tau=0.5$ lets \texttt{pyzenith} through,
Section~\ref{sec:results-discordance}). A properly powered,
cluster-aware test of the post-emission-registration effect over a
larger set of registered-but-suspicious packages is left to ongoing
work.

\paragraph{Construct validity.} Our ground-truth labels come from
PyPI live-probing at corpus-build time. PyPI is an evolving registry:
a package name unresolved today may resolve tomorrow if an attacker
or a legitimate maintainer registers it. We pin labels to the
probe-time snapshot and document the snapshot date in the
reproducibility material. Conversely, packages that existed at probe
time but are subsequently withdrawn under
PEP~592\footnote{\url{https://peps.python.org/pep-0592/}} remain
labelled as real. The metadata channel partially mitigates this
asymmetry: a withdrawn package retains its metadata profile (single
release, sparse author/summary) and the metadata channel re-routes
the flag to a posterior-confidence regime rather than an inappropriate
binary positive. Long-running deployments should re-probe periodically
and re-update the posterior with the freshly-observed registry state.

\paragraph{Internal validity --- Mahmud baseline fidelity.} The
Mahmud-shaped baseline is a single-stage binary detector inspired
by Mahmud et~al.~\cite{mahmud2025trustcalibrated}, not a verified
end-to-end replication of their full multi-stage trust-calibrated
pipeline. The fidelity gap is bounded above by
the trust-propagation stages we omit (which act on
already-flagged findings rather than on the strict-registry channel
itself); we quantify the gap when the original implementation is
publicly available.

\paragraph{Construct validity --- transitive dependency compromise.}
Our unit of analysis is the top-level import name emitted by the LLM.
A complementary attack --- a \emph{legitimate} package whose own
transitive dependencies, build scripts, or maintainer account have been
compromised --- is invisible to any import-time probe of the top-level
name, and we do not claim to address it (Section~\ref{sec:method-detector}).
That class of compromise is the remit of lock-file-level provenance and
integrity tooling (in-toto~\cite{torresarias2019intoto}, OSV-Scanner
over the resolved dependency graph) and of the empirical
supply-chain-attack measurements of Duan
et~al.~\cite{duan2021supplychain}; the calibration layer is
complementary to, not a substitute for, those defences.

\paragraph{External validity.} The corpus is Python-only. Python
hosts the most extensively documented slopsquat literature and
provides the BigCodeBench peer-reviewed corpus that our
realistic-use slice depends on. Slopsquat attacks generalise in
principle to other package ecosystems --- npm (whose security threats
Zimmermann et~al.~\cite{zimmermann2019npm} characterise), crates.io, Go
modules, RubyGems, and the Java/Maven Central ecosystem --- and the
underlying typosquatting vector was first measured across package
managers by Tschacher~\cite{tschacher2016typosquatting}. Our pipeline
is language-agnostic at the calibration layer, but the empirical
numbers reported here apply specifically to Python. We single out
Java/Maven Central as a particularly informative next ecosystem: its
group/artefact coordinate scheme and central-repository curation give a
\emph{different} registry-probe semantics from PyPI's flat namespace,
which would stress-test how much of the calibration transfers.
Cross-ecosystem generalisation is left to ongoing work.

\paragraph{LLM panel coverage.} Our six-model panel covers four cloud
models (Claude-Sonnet-4.6, Mistral-Large, DeepSeek-v4-pro, DeepSeek-R1)
and two local open-weight code models (Mistral Codestral, Meta
CodeLlama), spanning four developers (Anthropic, Mistral, DeepSeek,
Meta). Other closed cloud families (Google Gemini, OpenAI GPT-class
models, or Anthropic Opus-class models) are absent due to access and budget constraints at the time of
writing and remain a natural extension. The calibration layer is
panel-agnostic and the corpus can be re-scored with additional panel
models, or entirely a new set in future work.

\section{Reproducibility}
\label{sec:reproducibility}

We describe the experimental setup in sufficient detail for an
independent re-implementation. All pins below trace to artefacts
retained in the internal repository at corpus-build time
(2026-05-23) and to the inference logs.

\paragraph{Prompt source.} BigCodeBench v0.1.0\_hf (the canonical
Hugging Face split, $1140$ tasks)\footnote{\url{https://huggingface.co/datasets/bigcode/bigcodebench}}.
We stratify across the six library-use categories
(\texttt{http}, \texttt{ml}, \texttt{numerical}, \texttt{other},
\texttt{security}, \texttt{viz}) at \texttt{seed=20260520} and sample
$189$ prompts; the stratification reproduces deterministically
under the same seed. The adversarial complement is the $100$
prompts sampled from a $102$-prompt pool at
\texttt{seed=20260520}.

\paragraph{LLM panel (six models, pinned).}
\begin{itemize}[noitemsep,topsep=0pt]
  \item \texttt{claude-sonnet-4.6} via OpenRouter (route
    \texttt{anthropic/claude-sonnet-4.6}, accessed
    2026-05-20 -- 2026-05-22).
  \item \texttt{mistral-large-2411} via OpenRouter (route
    \texttt{mistralai/mistral-large-2411}, accessed
    2026-05-20 -- 2026-05-22).
  \item \texttt{deepseek-v4-pro} via the DeepSeek direct API
    endpoint at \url{https://api.deepseek.com/v1}, accessed
    2026-05-20 -- 2026-05-22.
  \item \texttt{deepseek-r1} via OpenRouter (route
    \texttt{deepseek/deepseek-r1}, accessed 2026-05-22).
  \item \texttt{codestral:22b}, local
    \texttt{ollama} digest \texttt{0898a8b286d5} (pull date 2026-05-16).
  \item \texttt{codellama:7b}, local \texttt{ollama} digest
    \texttt{8fdf8f752f6e} (pull date 2026-05-17).
\end{itemize}
Inference parameters are locked at $\text{temperature}=0.2$,
$\text{top\_p}=0.9$, $\text{max\_tokens}=2048$, $\text{variants}=1$.
Because temperature-$0.2$ sampling is not bit-reproducible across
cloud endpoints, reproducibility rests on the \emph{frozen corpus}:
the generated snippets are recorded once at corpus-build time (the
per-prompt completions in the inference logs above) and all results
trace to that saved corpus rather than to regenerating identical
completions.

\paragraph{PyPI probe snapshot.} The strict-registry ground-truth
label and the registry channel of variants~1--3 share a single
probe layer (\texttt{IMPORT\_TO\_DISTRIBUTION}-aware,
Section~\ref{sec:results-adv}) operating against the public PyPI
JSON endpoint at \url{https://pypi.org/pypi/<name>/json}. The
probe snapshot date is pinned to 2026-05-23; a re-probe at a
later date may
reclassify packages registered in the interim, an effect we
discuss in Section~\ref{sec:threats}. The 15-entry alias
whitelist is frozen at corpus-build time; extending it would
yield strictly $\geq$ the
ground-truth ``not-hallucinated'' set we report.

\paragraph{Implementation.} The strict-registry detector is a
$\sim$$100$-line Python script (AST extraction via
\texttt{ast-grep}, registry probe, Levenshtein-1 neighbour
search); the metadata channel adds another $\sim$$150$ lines to
read and route PyPI JSON metadata; the calibration layer is the
$\sim$$400$-line module described in
Section~\ref{sec:method-calibration}. The evaluation harness, the
full labelled corpus, and per-LLM raw outputs are retained
internally and may be made available to peer-review programme
committees on request, with broader release deferred to the
companion artefact submission. The cloud-routed models
(Claude-Sonnet-4.6 and Mistral-Large 2411 via OpenRouter;
DeepSeek-v4-pro via direct API) are reproducible from any account
with the equivalent routes; the two local-\texttt{ollama} models
are reproducible from the publicly published GGUF artefacts under
the digests above.

\section{Conclusion and perspectives}
\label{sec:conclusion}

We have presented a Bayesian calibration layer for slopsquat
detection that emits per-detection Beta-posterior probabilities with
explicit prior provenance, and extended it with a metadata channel
that surfaces \emph{registered-but-suspicious} packages corresponding
to the realistic post-LLM-emission attacker regime. Our empirical
evaluation on the full $N=1734$ corpus shows that (i)~modern
code-specialised LLMs on realistic-use prompts produce
strict-hallucination slopsquats at near-zero rate once import-name
aliases are accounted for; (ii)~the metadata channel introduces a
statistically significant discordance against the strict-registry
detector (McNemar $b=10$, $c=0$, $p=0.0020$), the empirical lever that
justifies calibration over a simple binary flag; and (iii)~the
calibrated confidences are well behaved, with flagged-subset ECE
($0.016$ / $0.029$) and a strictly proper full-corpus Brier score
($\approx 0.022$) both well inside conventional thresholds across
detection channels.

Several directions extend this work naturally: cross-ecosystem
generalisation to npm, crates.io, Go module and Java/Maven Central
registries; an end-to-end pipeline coupling detection with remediation
and provenance attestation; probabilistic verification of the
cat.~3~$\to$~cat.~2~$\to$~cat.~1 promotion machine
(Section~\ref{sec:method-taxonomy}) as a model-checking property; and
adversarial robustness studies of calibration under adaptive priors. We
see the calibration primitive as a building block for risk-aware
AI-assisted software supply-chain defence, complementary to detection
mechanisms targeting other
classes of LLM-induced supply-chain risk.

\subsubsection*{Acknowledgements}

This work was carried out as part of the NewCo Partners research
programme on software quality and security for AI-assisted software
development. The calibrated detection primitive presented here is one
contribution to that programme.

\bibliographystyle{plain}
\bibliography{bibliography}

\end{document}